\documentclass{vietnam}
\usepackage{graphicx, subfigure}

\def\be{\begin{equation}}
\def\ee{\end{equation}}
\def\bea{\begin{eqnarray}}
\def\eea{\end{eqnarray}}

\begin{document}
 \vspace*{4cm}
\begin{center}
{\Large \bf{Full $\mathcal{O}(\alpha)$ electroweak radiative corrections to
$t \bar{t} \gamma$ and $e^-e^+\gamma$ productions at ILC with GRACE-Loop}}\\
 \vspace*{0.3cm}
\small{P.H. Khiem$^{A,B}$, Y. Kurihara$^{A}$,  J. Fujimoto$^{A}$, M. Igarashi$^{C}$, T. Ishikawa$^{A}$,
T. Kaneko$^{A}$, K. Kato$^D$,\\
N. Nakazawa$^{D}$, Y. Shimizu$^A$, K. Tobimatsu$^{D}$, 
T. Ueda$^{E}$, J.A.M. Vermaseren$^F$, Y. Yasui$^{G}$} \\
 \vspace*{0.3cm}
\textit{\small{$^{A)}$KEK, Oho 1-1, Tsukuba, Ibaraki 305-0801, Japan.\\
  $^{B)}$SOKENDAI University, Shonan Village, Hayama, Kanagawa 240-0193 Japan. \\ 
  $^{C)}$Tokai University, Hiratsuka-shi, Kanagawa, 259-1292 Japan.\\
  $^{D)}$Kogakuin University, Shinjuku, Tokyo 163-8677, Japan.\\ 
  $^{E)}$Karlsruhe Institute of Technology (KIT), D-76128 Karlsruhe, Germany. \\
  $^{F)}$NIKHEF, Science Park 105, 1098 XG Amsterdam, The Netherlands.\\ 
  $^{G)}$Tokyo Management College, Ichikawa, Chiba 272-0001, Japan.}
  }
\end{center}
\abstracts{
The full  $\mathcal{O}(\alpha)$ electroweak radiative corrections to
$t \bar{t} \gamma$ and $e^-e^+\gamma$ production at International Linear Collider (ILC) are presented in this paper.
The computation is performed with the help of GRACE-Loop system. In the physical results, we discuss on the cross section,
electroweak corrections, and the top quark forward-backward asymmetry ($A_{FB}$) which are the function of the 
center-of-mass energy. 
}

\section{Introduction}
The main goals of future colliders, such as ILC, are not only to make precise measurements Higgs properties as well as 
top quarks, vector bosons interactions, but also to search for physics beyond the Standard Model (BSM). At ILC 
the measurements will be performed
with high precision, with the statistical error below $0.1\%$ typically.  Therefore, when we talk about
the processes at $e^-e^+$ collision the  full one-loop electroweak radiative 
corrections should be important. In this paper, we present full $\mathcal{O}(\alpha)$ 
electroweak radiative corrections to the most important two reactions, 
$e^- e^+ \rightarrow t \bar{t} \gamma$ and $e^- e^+ \rightarrow e^-e^+\gamma$. 
In the numerical analysis, we examine the $A_{FB}$ of
$t \bar{t} \gamma$ production as well as the corrections to the total cross section of the reactions.

The paper is organized as follow. In the section 2, we introduce the GRACE-Loop system and explain how to make the check 
of the calculations. Then the physical results of the calculations will be discussed in the section 3. The section 4 is devoted
to the conclusion and future plan of the paper. 

\section{GRACE-Loop and test of the calculation}
\subsection{GRACE-Loop}
GRACE-Loop is a generic program for the automatic calculation of scattering processes 
in High Energy Physics~\cite{Belanger:2003sd}. In the system, the renormalization has been carried out with
the on-shell renormalization conditions of the Kyoto scheme~\cite{kyotorc}. The GRACE-Loop have also been implemented non-linear gauge fixing 
conditions~\cite{nlg-generalised}. With the results being independent non-linear gauge parameters, 
the ultraviolet and the infrared finiteness, the system provides a power 
tool to check the calculation as discussed in detail in the next section. 

Recently, we have implemented the axial gauge for external photon into the GRACE-Loop which brings two advantages. 
First it cures a problem with large numerical cancellation. 
This is very useful when we calculate the processes at small angle and energy cuts for the particles in the final state.
Secondly it provides with a useful tool to check the consistency of the results which  should be
independent of the choice of the gauge.
\subsection{Test of the calculation}
In this subsection, we discuss on the test of the calculation in detail. We take the process 
$e^- e^+ \rightarrow t \bar{t} \gamma$ as an example. The corrected total cross section is composed of the tree 
and the full one-loop graphs together with the 
soft and hard bremsstrahlung contributions. In general the total cross section should be independent of
ultraviolet cutoff parameter ($C_{UV}$), the fictitious photon 
mass ($\lambda$) as well as the minimum cut of hard photon energy ($k_c$) and five gauge parameters 
($\tilde{\alpha}, \tilde{\beta}, \tilde{\kappa},\tilde{\delta}, \tilde{\epsilon}$)\cite{nlg-generalised}.
The table~(\ref{cuv-gauge}) shows the test for $C_{UV}$, $\lambda$, and gauge 
parameters independence of the amplitude at one phase space point arbitrarily chosen. We find that the results are stable 
over $19$ digits. In the table~(\ref{kc}), the $k_c$-stability of the result is presented. This test is done
at the cross section level,  we find that the results are in agreement with an accuracy better than $0.1\%$ when we vary $k_c$.

\begin{table}[h]
\caption{\label{cuv-gauge} The $C_{UV}$, $\lambda$ and gauge parameters independence of the amplitude 
at $1$ TeV center-of-mass energy. In this table, we set $k_c = 10^{-3}$ GeV.}
\centering
\begin{tabular}{lc} \hline
$(C_{UV}, \lambda,\{\tilde{\alpha}, \tilde{\beta}, \tilde{\kappa},\tilde{\delta}, \tilde{\epsilon} \}) $ 
& $2\mathcal{\Re}(\mathcal{T}_{Tree}^{+}\mathcal{T}_{Loop})+$ soft contribution\\ \hline  \hline 
$(0,10^{-21},\{0,0,0,0,0\})$ & $-2.5673606057317280692041118561248255\cdot 10^{-3} $\\ \hline
$(10,10^{-22},\{1,2,3,4,5\})$ &      $-2.5673606057317280690196417213681083\cdot 10^{-3} $\\ \hline
\end{tabular}
\end{table}
\vspace*{-0.4cm}
\begin{table}[h]
\caption{\label{kc} Test of the $k_c$-stability of the result. We 
choose the photon mass to be $10^{-17}$ GeV and the center-of-mass energy 
is $1$ TeV. The second column presents the hard photon cross-section and 
the third column presents the soft photon cross-section. The final column 
is the sum of both.}
\centering
\begin{tabular}{lccc} \hline 
 $k_c $[GeV] & $ \sigma_{H}$ [pb]  & $ \sigma_{S}$ [pb] &  $ \sigma_{S+H}$ [pb] \\ \hline  \hline
$10^{-3} $ & $2.92668\cdot 10^{-02} $  & $7.13173\cdot 10^{-02} $ &  $0.100584$ \\ \hline    
$10^{-1} $ & $1.67899\cdot 10^{-02} $  & $8.37731\cdot 10^{-02} $ &  $0.100563$ \\ \hline   
\end{tabular}
\end{table}

\section{The physical results}
In this section, we will present the numerical 
analysis of $t\bar{t}\gamma$ and $e^+e^-\gamma$ productions at ILC.
Our input parameters for the calculation are as follows. The fine structure 
constant in the Thomson limit is $ \alpha^{-1}=137.0359895$. The mass of 
the Z boson is $M_Z=91.187$ GeV, the mass of W boson is $M_W=80.3759$ GeV and 
$M_H=120$ GeV. For the lepton masses we take $m_e=0.51099891$ MeV, 
$m_{\tau}=1776.82$ MeV and $m_{\mu}=105.658367$ MeV. For the quark 
masses we take $m_u=1.7$ MeV, $m_d=4.1$ MeV, $m_c=1.27$ GeV, $m_s=101$ MeV,
$m_t=172.0$ GeV and $m_b=4.19$ GeV. 

\subsection{The process $e^- e^+ \rightarrow t \bar{t} \gamma$}
\begin{figure}[ht!]
     \begin{center}
        \subfigure[]{%
            \label{a}
            \includegraphics[width=2in,height=7.5cm,angle=-90]{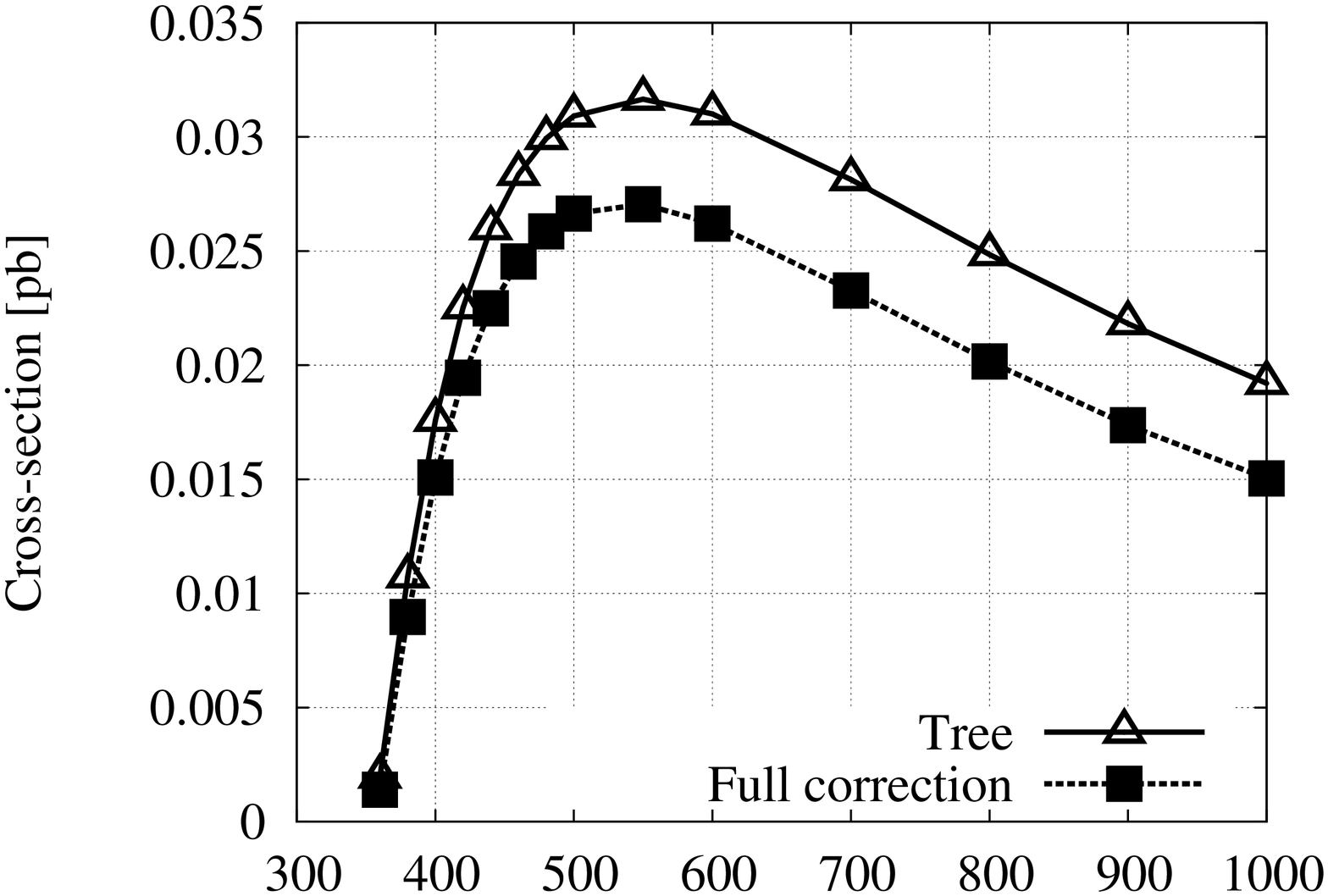}
        }%
        \subfigure[]{%
           \label{b}
           \includegraphics[width=2in,height=7.5cm,angle=-90]{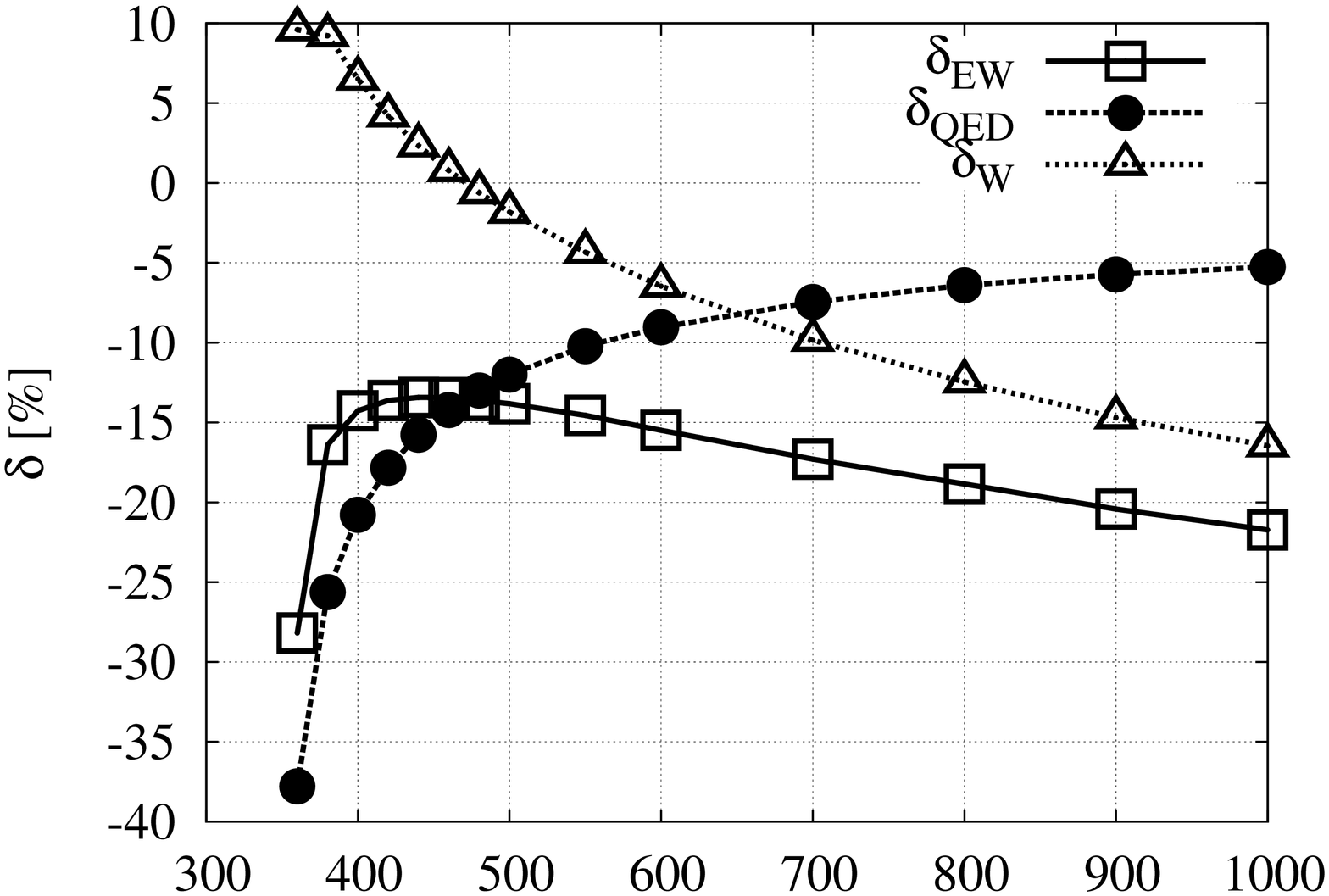}
        }\\ 
        \vspace*{-0.5cm}
        \subfigure[]{%
            \label{c}
            \hspace*{0.3cm} \includegraphics[width=2.in,height=7.2cm,angle=270]{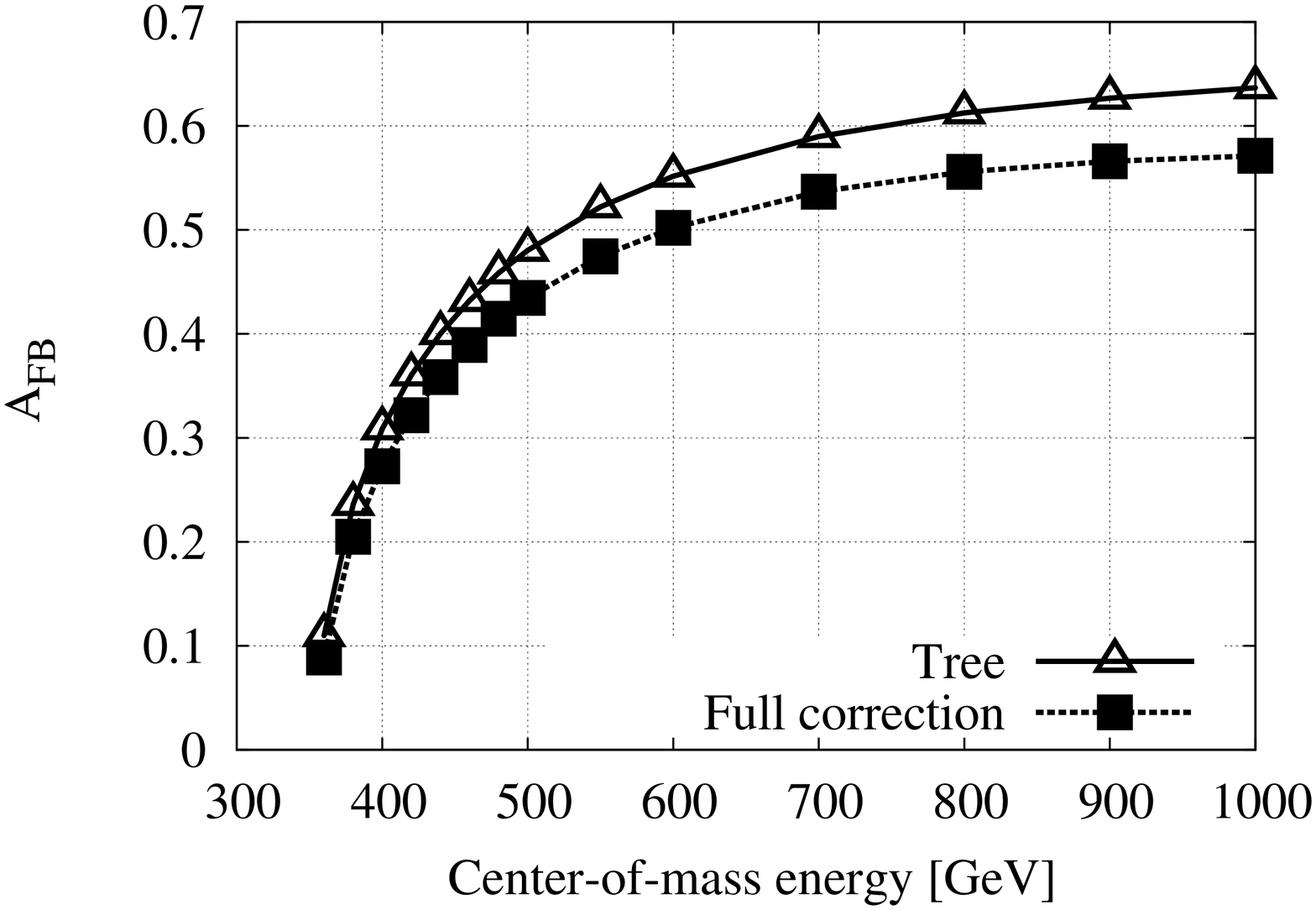} 
        }%
        \subfigure[]{%
            \label{d}
             \hspace*{-0.5cm}\includegraphics[width=2.in,height=7.8cm,angle=270]{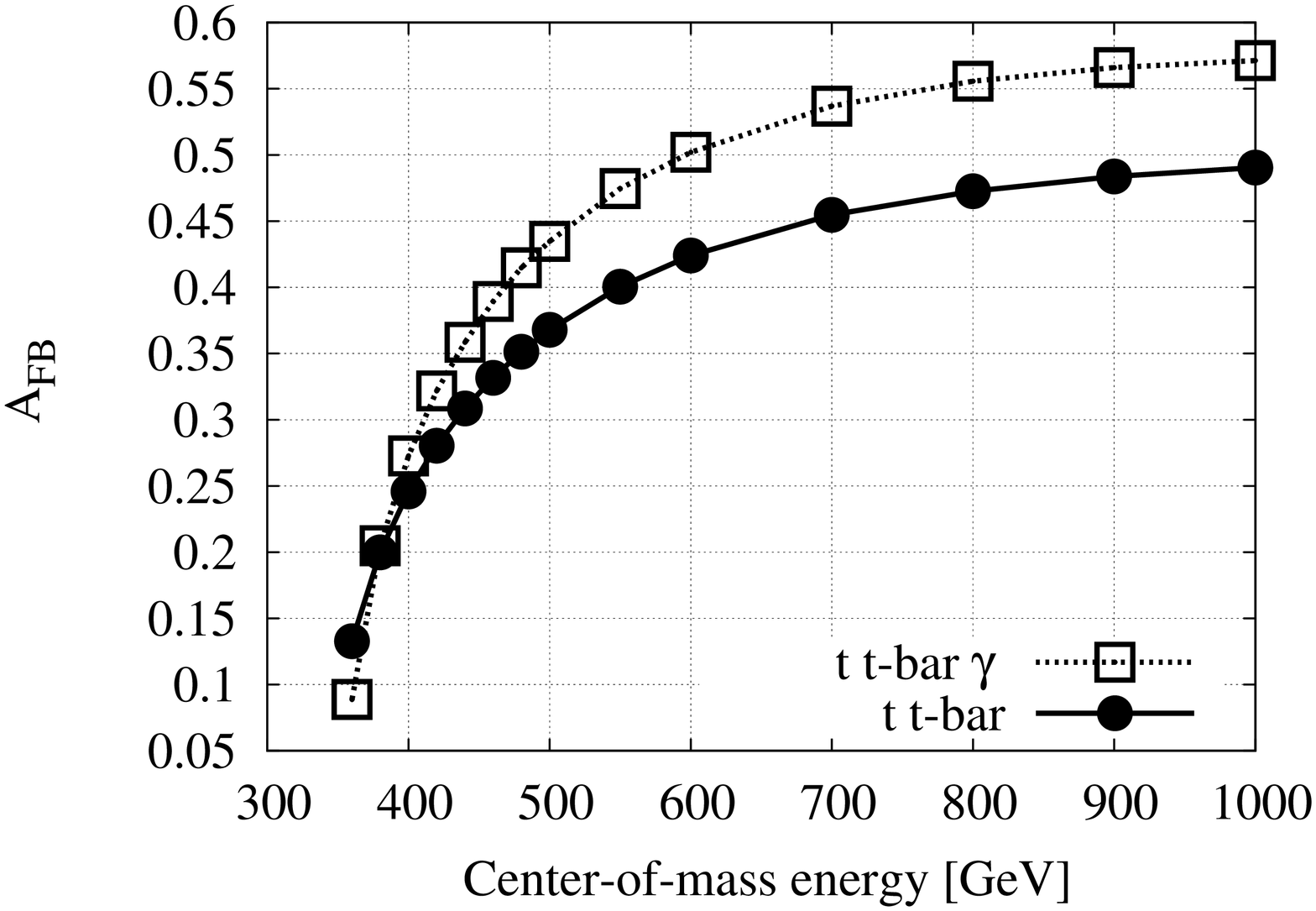} 
        }%
    \end{center}
    \caption{The total cross-section, the full electroweak corrections as well as the 
genuine weak correction in the $\alpha$ scheme and $A_{FB}$, which are shown as a 
function of the center-of-mass energy.}
   \label{abcd}
\end{figure}

For this process, we impose an energy cut $E_{\gamma}^{cut} \geq 10$ GeV and 
an angle cut $10^{\circ}\leq \theta_{\gamma}^{cut} \leq 170^{\circ}$ on the photon. 
In Fig~(\ref{abcd}) the total cross-section, the full electroweak corrections as well as the 
genuine weak correction in the $\alpha$ scheme and $A_{FB}$, which are shown as a 
function of the center-of-mass energy ($\sqrt{s}$),  are presented. 
We vary $\sqrt{s}$ from $360$ GeV to $1$ TeV. We find that the cross-section is largest near the 
threshold, $\sqrt{s}$ around $550$ GeV and it decreases when $\sqrt{s}$ increases. The Fig~\ref{b} shows 
clearly that the QED corrections is dominant in the low energy region. 
In the high energy region it is getting much smaller. In contrast to the QED corrections 
the weak corrections in the $\alpha-$scheme is less than $10\%$ in low energy but reaches 
$-16\%$ at $1$ TeV. In the
Fig~\ref{c}, we find that the top quark 
asymmetry in the full correction is smaller than that one at the tree 
level. Therefore, the electroweak corrections to this process
plays important role for $A_{FB}$ measurement in the future collider. In the last Fig~\ref{d}, 
we compare the $A_{FB}$ in $t\bar{t}\gamma$ to that one in $t\bar{t}$~\cite{Khiem:2012bp}. The 
former is greater than the latter case  when $\sqrt{s}\geq 400$GeV.
This must be clearly observed at ILC.

\subsection{The process $e^- e^+ \rightarrow e^- e^+ \gamma$}

\begin{figure}[ht!]
     \begin{center}
        \subfigure[]{%
            \label{a2}
            \hspace*{-0.5cm}\includegraphics[width=2.in,height=8.5cm,angle=-90]{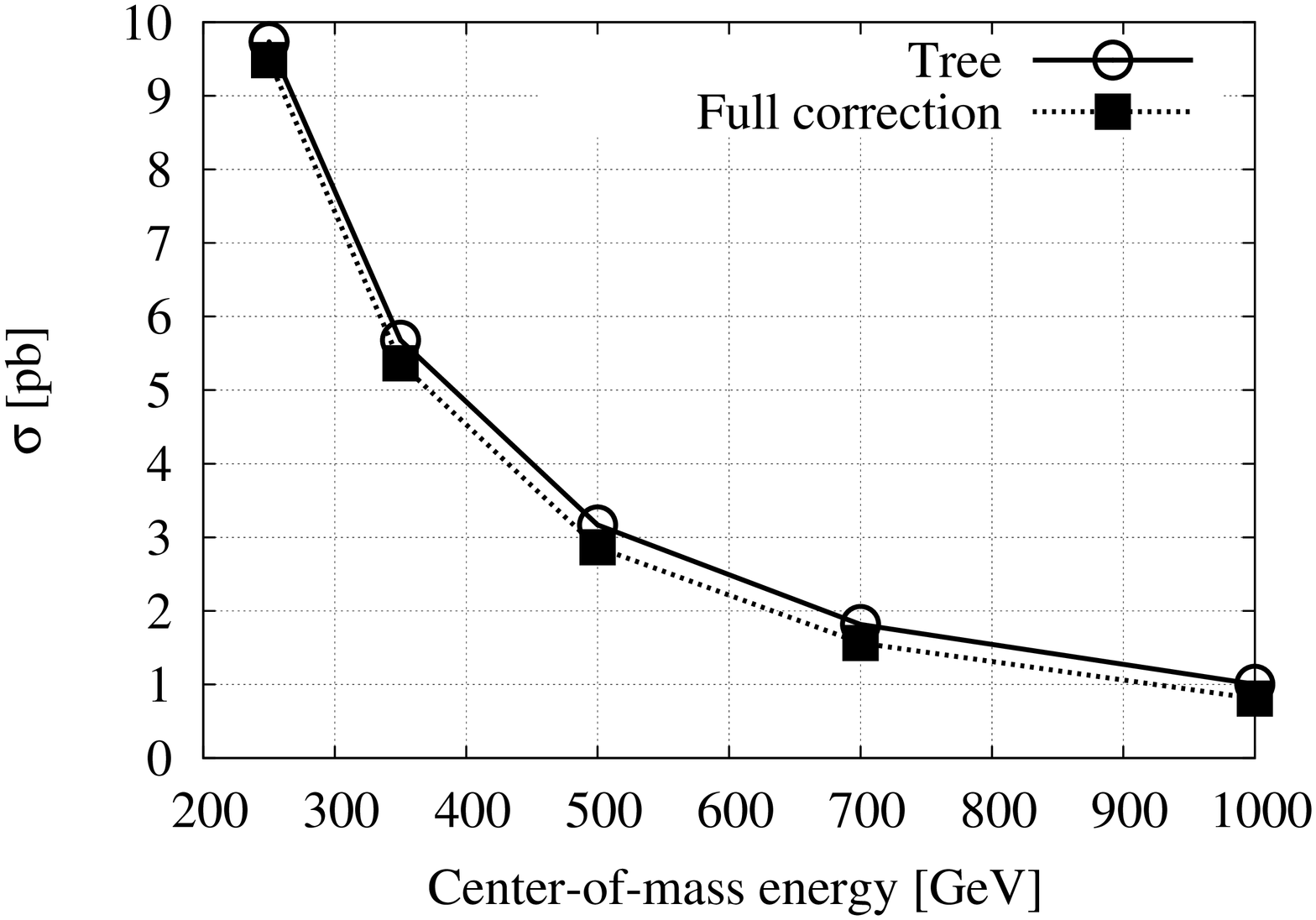}
        }%
        \subfigure[]{%
           \label{b2}
            \hspace*{-1.cm} \includegraphics[width=2.in,height=9.5cm,angle=-90]{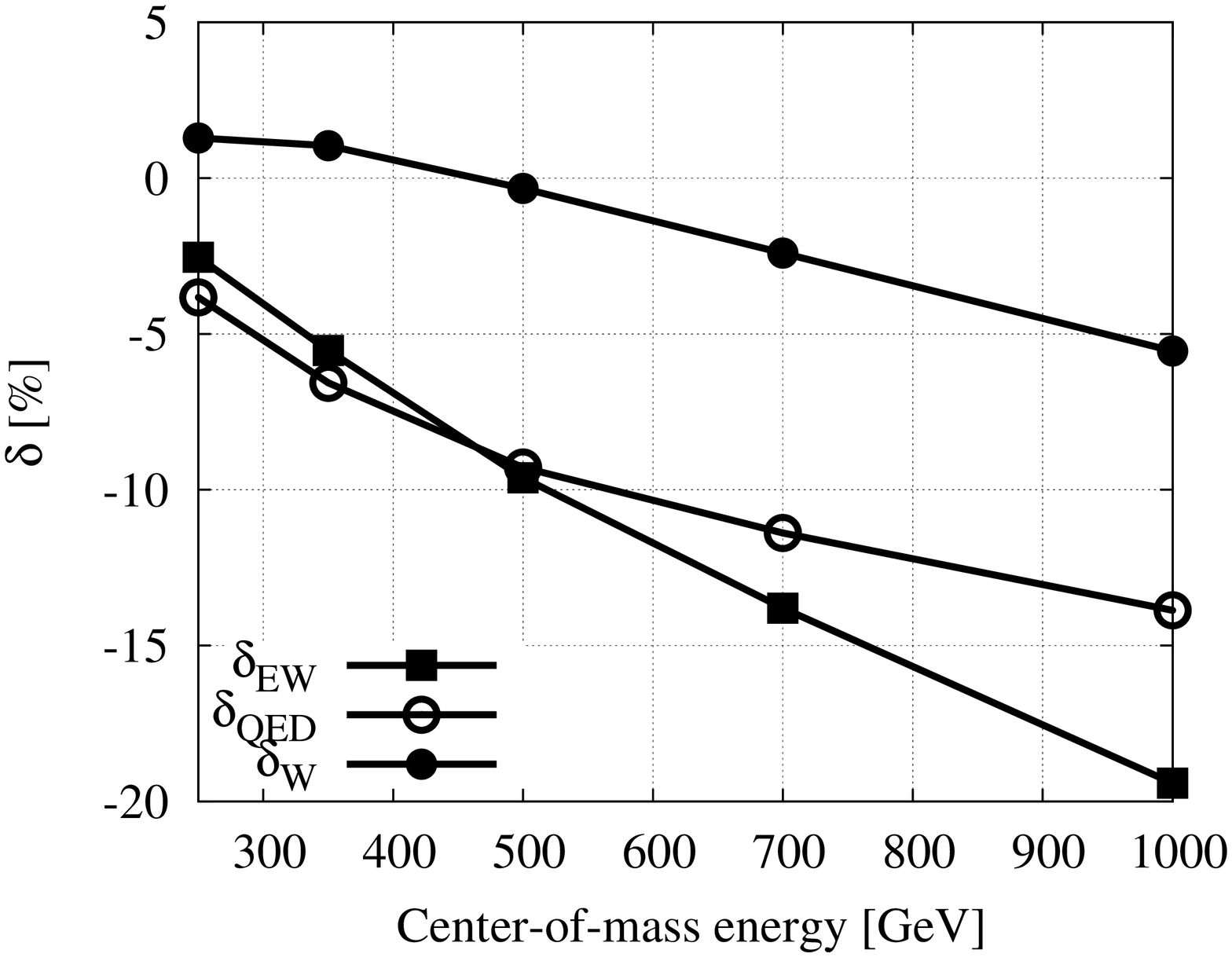}
       
        }\\ 
             \vspace*{-0.5cm}
        \subfigure[]{%
             \label{c2}
             \hspace*{-1.85cm} \includegraphics[width=2.5in,height=10.5cm,angle=270]{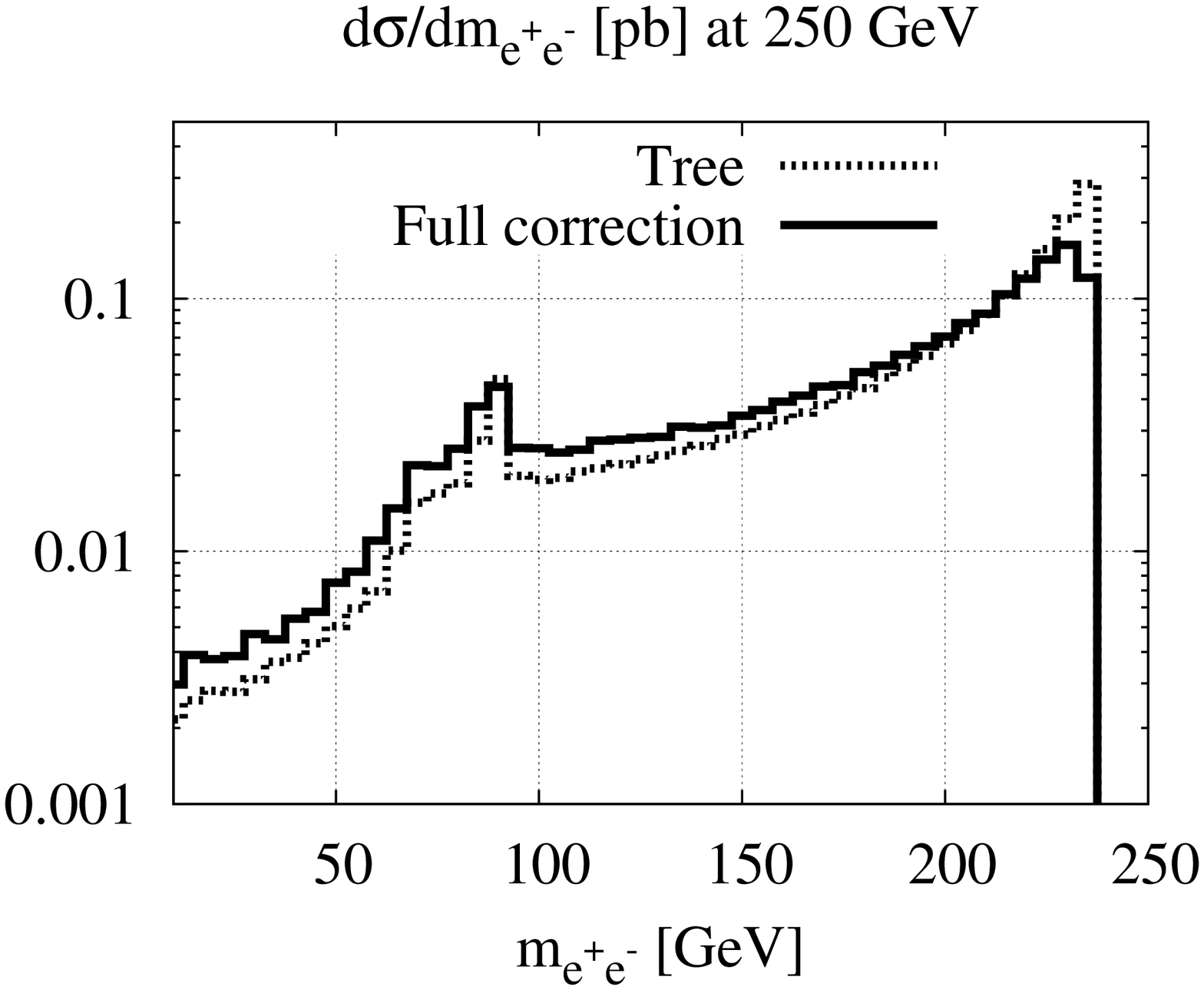}
        }%
        \subfigure[]{%
            \label{d2}
           \hspace*{-2.75cm} \includegraphics[width=2.5in,height=10.5cm,angle=270]{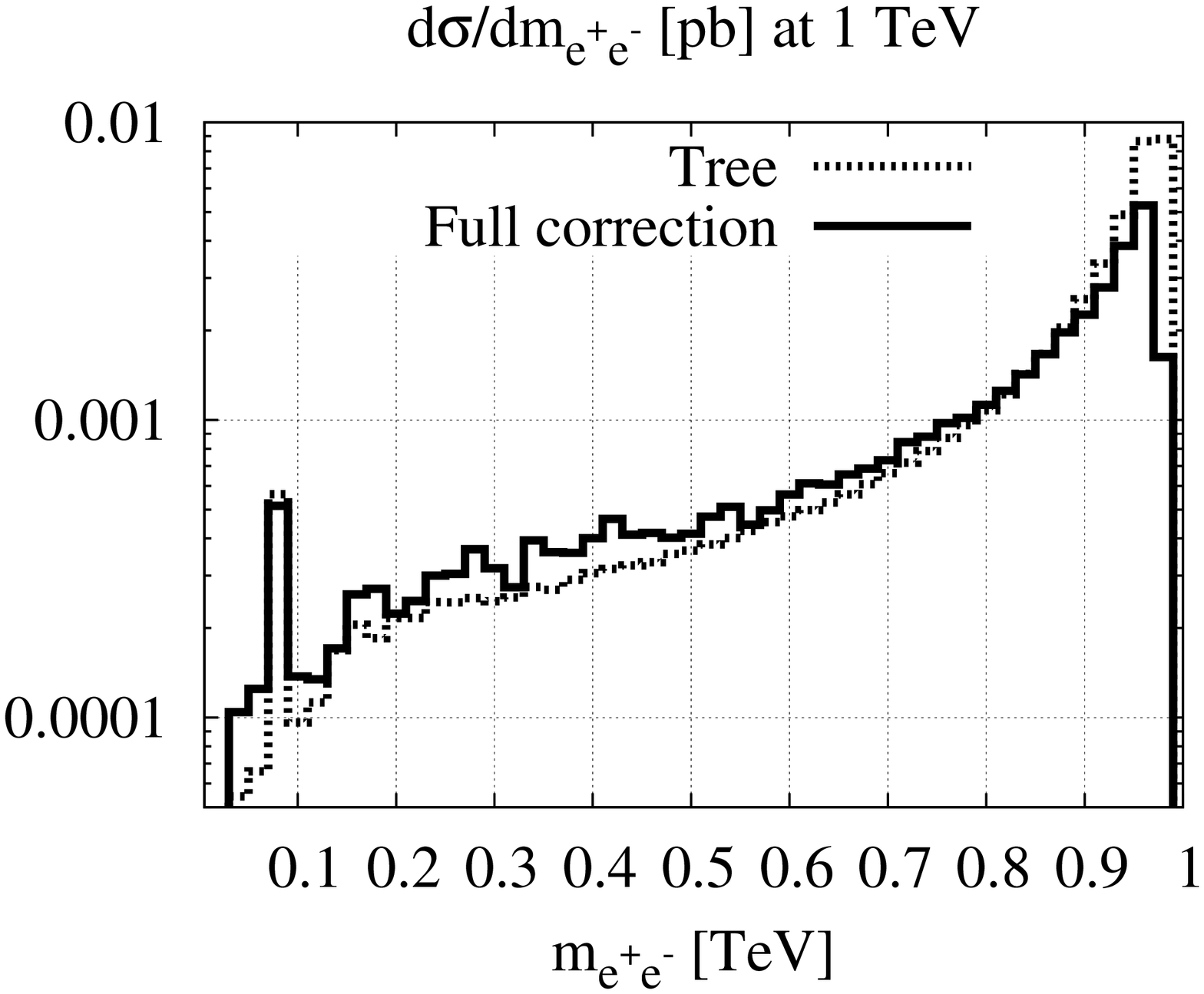} 
        }%
\end{center}
\caption{The cross-section, the electroweak corrections which are
as a function of $\sqrt{s}$ and 
the invariant mass of $e^-$, $e^+$
distribution at $\sqrt{s}=250$ GeV ($\sqrt{s}=1$ TeV) are shown. }
   \label{abcd2}
\end{figure}

For the calculation of corrections for this process we impose two cuts on the final state particles: an energy cut $E_\gamma^{cut}\ge10$ 
Gev and an angle cut $10^\circ\le\theta_\gamma^{cut}\le170^\circ$ off the beam line. Further in order to isolate the photon from 
the electron (and positron) we require the open angle cut between the photon and $e^- (e^+)$ to be $10^\circ$. Also the angle between
the electron and the positron should be greater than $10^\circ$. In Fig~(\ref{abcd2}) the cross section and the electroweak corrections as functions of $\sqrt{s}$ and also the invariant mass
distributions of $e^-,e^+$ at $\sqrt{s}=250$ GeV  and $\sqrt{s}=1$ TeV, are shown. The center-of-mass energy ranges from 250 GeV
(the threshold of $M_Z+M_H$) to 1 TeV. The cross  section decreases in the high energy region. The electroweak corrections are
found to decrease from $-2\%$ to $-20\%$ when $\sqrt{s}$ varies from 250 GeV to 1 TeV. In Figs. 2(c) and 2(d) peaks appear
near $M_Z$ and high mass side (it corresponds to radiative tail). From these distributions we see that the radiative corrections are
visible and thus it should be important for luminosity monitor of ILC.
\vspace*{-0.1cm}
\section{Conclusion}
We have presented the full $\mathcal{O}(\alpha)$ electroweak 
radiative corrections to the process $e^+e^- \rightarrow t \bar{t} \gamma$ and
$e^+e^- \rightarrow e^+e^-\gamma$
at ILC. The calculations were done with the help of 
the GRACE-Loop.  Concerning the top quark productions we found that the 
corrections give sizable contribution to $A_{FB}$ as well as the total cross section. 
Also the corrections to the reactions related to the
Bhabha scattering are very important for high precision measurement and thus for luminosity monitoring at ILC. 
\section*{Acknowledgments}
We wish to thank Dr. F.~Yuasa and Dr. N.~Watanabe for 
their valuable discussions and comments. This work was supported by 
Grant-in-Aid for Scientific Research (No.24540292) of JSPS

\end{document}